\documentclass[10pt]{article}

\usepackage[superscript,sort]{cite}

\usepackage{ifthen,ifpdf}
\ifpdf
	\usepackage[pdftex]{hyperref}	\hypersetup{colorlinks=true,linkcolor=black,citecolor=black,urlcolor=blue,backref=page,bookmarks=true,breaklinks=true,plainpages=false }
	\usepackage[pdftex]{graphicx}
	\usepackage[usenames,pdftex]{color}
\else
	\usepackage[colorlinks=true,breaklinks=true,plainpages=false]{hyperref}
	\usepackage{graphicx}
	\usepackage[usenames]{color}
\fi

\usepackage{amsthm,amscd,amsxtra,amsfonts,amsmath,amssymb,multirow}
\usepackage{wrapfig}
\usepackage[footnotesize]{caption}
\usepackage[tiny,compact]{titlesec}
\usepackage[textwidth=0.8in,textsize=footnotesize]{todonotes}
\usepackage{algorithm,algorithmic,extarrows}

\begin{document}

\title{Persistent homology analysis of ion aggregation and hydrogen-bonding network}

\author{
Kelin Xia$^1$ \footnote{ Address correspondences  to Kelin Xia. E-mail:xiakelin@ntu.edu.sg} \\
$^1$Division of Mathematical Sciences, School of Physical and Mathematical Sciences, \\
Nanyang Technological University, Singapore 637371\\
}

\date{\today}
\maketitle

\begin{abstract}
Despite the great advancement of experimental tools and theoretical models, a quantitative characterization of the microscopic structures of ion aggregates and its associated water hydrogen-bonding networks still remains a challenging problem.
In this paper, a newly-invented mathematical method called persistent homology is introduced, for the first time, to quantitatively analyze the intrinsic topological properties of ion aggregation systems and hydrogen-bonding networks.
Two most distinguishable properties of persistent homology analysis of assembly systems are as follows. First, it does not require a predefined bond length to construct the ion or hydrogen network. Persistent homology results are determined by the morphological structure of the data only. Second, it can directly measure the size of circles or holes in ion aggregates and hydrogen-bonding networks. To validate our model, we consider two well-studied systems, i.e., NaCl and KSCN solutions, generated from molecular dynamics simulations. They are believed to represent two morphological types of aggregation, i.e., local clusters and extended ion network. It has been found that the two aggregation types have distinguishable topological features and can be characterized by our topological model very well. Further, we construct two types of networks, i.e., O-Networks and ${\rm H_2O}$-Networks, for analyzing the topological properties of hydrogen-bonding networks. It is found that for both models, KSCN systems demonstrate much more dramatic variations in their local circle structures with the concentration increase. A consistent increase of large-sized local circle structures is observed and the sizes of these circles become more and more diverse. In contrast, NaCl systems show no obvious increase of large-sized circles. Instead a consistent decline of the average size of circle structures is observed and the sizes of these circles become more and more uniformed with the concentration increase. As far as we know, these unique essential topological features in ion aggregation systems have never been pointed out before. More importantly, our models can be directly used to quantitatively analyze the intrinsic topological invariants, including circles, loops, holes, and cavities, of any network-like structures, such as nanomaterials, colloidal systems, biomolecular assemblies, among others. These topological invariants cannot be described by traditional graph and network models.
\end{abstract}

Key words: persistent homology, ion aggregation, hydrogen-bonding network, Betti number, Barcodes, simplicial complex



\newpage

\section{Introduction}

Water and ion play irreplaceable roles in life and living organisms. Their unique properties are highly related to hydrogen-bonding network and ion aggregation.
It is known that water molecules can form strong and extensive hydrogen bond network\cite{Marcus:2009effect}. Based on different criteria, it has been estimated that each water molecule on average has about 2.8 to 3.4 hydrogen bonds \cite{Xenides:2006hydrogen, kumar:2007hydrogen}. The hydrogen-bonding network is a gigantic and dynamic structure made from hydrogen bonds, which can form and break incessantly on picosecond timescale\cite{laage:2011water}.
On the other hand, the ion is one of the most important solutes. They can easily dissolve into water and form ion aggregates, which usually have complicated morphological structures. Moreover, different ions have dramatically different properties, such as viscosity, heat capacity, and activity coefficient\cite{Lo:2012hofmeister}. This is known as ion specificity or Hofmeister effects\cite{zhang:2006interactions,Marcus:2009effect,Lo:2012hofmeister}. It is believed that ion specificity is highly related to`` structure making" and ``structure breaking" of hydrogen networks\cite{zhang:2006interactions}. Chaotropic ions, such as ${\rm SCN}^-$ and ${\rm ClO_4}^-$, are capable of breaking the water hydrogen bonds, whereas kosmotropic ions, such as ${\rm CO_3}^{2-}$ and ${\rm SO_4}^{2-}$ can make or stabilize water hydrogen bonds. Extensive experiments have been done to reveal the microscopic structure of ion aggregates and water hydrogen-bonding network\cite{cerreta:1987structure,rusli:1989raman,bian2011ion,dixit2002molecular,dillon:2003nmr,mason:2004structure,mason:2005nanometer,Wallace:2013microscopic,Shiraga:2016hydrogen}. Only recently, it has been found that ions had no obvious influence on hydrogen bonds outside the first solvation shell\cite{Omta:2003negligible}. However, a quantitative characterization of the complicated morphological structures and dynamics of ion aggregates and hydrogen-bonding networks in atomic scale is still a challenging problem \cite{Marcus:2009effect}.

Currently, the most promising approach to reveal atom-level structures of ion aggregates and hydrogen-bonding networks is an extensive comparison between available experimental observables and computational simulations\cite{mason:2004structure,mason:2005nanometer,Xenides:2006hydrogen,kumar:2007hydrogen}. 
However, with morphological complexity in ion aggregates and hydrogen networks, the available mathematical methods, that can quantitatively analyze the simulation data, is very limited. Historically, graph and/or network-based models are the dominant tools for ion aggregation and hydrogen-bonding network analysis\cite{radhakrishnan:1991graph, dos:2004topology,Oleinikova:2005formation}. Especially, the spectral graph theory has been widely used in chemistry and biochemistry to characterize the network properties\cite{Bako:2008water,da:2011hydrogen,Bako:2013hydrogen}, such as degree distribution, path length, clustering, among others.
These graph-based models and descriptors have greatly enhanced our understanding of ion and water aggregation systems. Recently, a new topological method, called persistent homology, has demonstrated its exceptional ability in revealing the intrinsic topological information of structures \cite{Edelsbrunner:2002,Zomorodian:2005,Zomorodian:2008}. Dramatically different all the graph or network models, persistent homology provides a way to quantitatively characterize topological invariants, such as connected components, circles, rings, loops, channels, cavities, voids, etc, in the data. It brings a whole new perspective for ion aggregation and hydrogen-bonding network analysis.

As a multiscale representation of topological features, persistent homology is able to bridge the gap between geometry and topology. Compared with traditional topological models, it manages to incorporate a geometric measurement into topological invariants through a process called filtration\cite{edelsbrunner:2010,Kaczynski:2004}. More specifically, by varying the value of a filtration parameter, a series of simplicial complexes are generated. These nested simplicial complexes encode topological information of a structure from different scales. Some topological invariants ``live" constantly in these simplicial complexes, whereas others disappear very quickly when filtration value changes. In this way, topological invariants can be quantified by their ``lifespans" or ``persisting times", which are directly related to geometric properties\cite{Dey:2008,Dey:2013,Mischaikow:2013}. Persistent homology has already been successfully used in a variety of fields, including shape recognition \cite{DiFabio:2011}, network structure \cite{Silva:2005,LeeH:2012,Horak:2009}, image analysis \cite{Carlsson:2008,Pachauri:2011,Singh:2008,Bendich:2010,Frosini:2013}, data analysis \cite{Carlsson:2009,Niyogi:2011,BeiWang:2011,Rieck:2012,XuLiu:2012}, chaotic dynamics verification \cite{Mischaikow:1999}, computer vision \cite{Singh:2008}, computational biology \cite{Kasson:2007,YaoY:2009, Gameiro:2013}, amorphous material structures\cite{hiraoka:2016hierarchical,saadatfar:2017pore}, etc. Various softwares, including JavaPlex \cite{javaPlex}, Perseus  \cite{Perseus}, Dipha \cite{Dipha}, Dionysus \cite{Dionysus}, jHoles \cite{Binchi:2014jholes}, GUDHI\cite{gudhi:FilteredComplexes}, etc, have been developed\cite{fasy:2014introduction}. The results from persistent homology can be visualized by many methods, including persistent diagram\cite{Mischaikow:2013}, persistent barcode\cite{Ghrist:2008barcodes}, and persistent landscape\cite{Bubenik:2007,bubenik:2015}. More recently, persistent homology has been used in analyzing biomolecular structure, flexibility and dynamics\cite{KLXia:2014c, KLXia:2015a,BaoWang:2016a}. The consistent pattern within the barcode is defined as topological fingerprint, which is used to quantitatively characterize biomolecular conformational information. Further, multiresolution and multidimensional persistent homology \cite{KLXia:2015c,KLXia:2015b} have been designed to extract topological information from various scales and have been used in handling extremely large data from macroproteins or protein assembly.
Most recently, the combination of persistent homology and machine learning tools has generated remarkable results in drug design\cite{cang:2017topologynet,cang:2017integration,nguyen:2017rigidity}, protein stability changes upon mutation\cite{cang:2017analysis,cang:2018representability} and  toxicity prediction\cite{wu:2018quantitative}. The great success of these models demonstrates that specially-designed persistent homology models can not only retain critical chemical and biological information during the simplification of biomolecular complexity, and also enables a tailored topological description of inter- and/or intra-molecular interactions of interest\cite{cang:2018representability}.

In this paper, persistent homology is introduced, for the first time, to quantitatively analyze the morphological structure of ion aggregates and hydrogen-bonding networks. We consider two well-studied systems, NaCl and KSCN solutions, as test examples\cite{Kim:2014ionI,Choi:2014ionII}. It has been found that when the ion concentration is close to their solubility limits, these two systems have two types of morphologically different ion aggregations\cite{Kim:2014ionI,Choi:2014ionII,choi2016ion}. For NaCl system, large compact ion clusters are formed and segregated from the water. In contrast,
KSCN develops spatially extended ion network structure, which tightly intertwines with water hydrogen-bonding network. Moreover, the KSCN network is found to bear great resemblance with hydrogen-bonding network\cite{Choi:2014ionII}. In our persistent homology analysis, barcodes information has been generated from both systems in various concentrations, it is found that when the concentration of NaCl is close to their solubility limits, a group of barcodes appear at filtration size around 3.0 \AA~ to 4.0 \AA,~ an indication of compact ion clusters. Further, when KSCN is close to solubility limits, a gigantic amount of barcodes appear in the region from 3.0 \AA~ to 4.0 \AA,~ a clear manifestation of extended ion networks. Moreover, barcode properties of KSCN ion aggregation network are quite similar to those from hydrogen-bonding network in high KSCN concentration. Our results are highly consistent with previous findings\cite{Kim:2014ionI,Choi:2014ionII}. More interesting, we find that similar cluster structures, as in the case of NaCl solution with concentration close to solubility limits, appear in KSCN solution with mediate concentration. This has never been pointed out before.

Further, we consider the hydrogen-bonding networks by analyzing two sets of points cloud data. The first one is made of only oxygen atoms, and it is called O-Network. The other one comprises oxygen and hydrogen atoms, and it is denoted as ${\rm H_2O}$-Networks. From the barcodes results and measurements, we find that, topologically, the fundamental difference between the two types of ion aggregation can be well characterized by the number, mean and variance of circles in their networks. More specifically, in O-Networks, with the increase of ion concentration, total number of local circle structures in NaCl and KSCN all decline, whereas KSCN shows a much larger decreasing rate. Moreover, both mean and variance of local circle sizes for NaCl decrease with concentration, whereas the circle mean and variance for KSCN have exactly the opposite behavior. In ${\rm H_2O}$-Networks, the number of NaCl circles increase with the concentration, whereas this number for KSCN decreases. Finally, both NaCl and KSCN systems show a decrease in their average circle size. Their behavior in size variance is exactly the opposite. With concentration increase, NaCl variance decreases with it, but KSCN variance increases.

The paper is organized as follows, Section \ref{sec:method} is devoted for persistent homology theory and models. We discuss the basic concepts, algorithms, and models of persistent homology. The detailed results of topological analysis of NaCl and KSCN systems are given in Section \ref{sec:results}. The paper ends with a conclusion.

\section{Theory and Models} \label{sec:method}

This section is devoted to the basic theory and models of persistent homology. In the first part, we present a brief review of key concepts in persistent homology, including simplex, simplicial complex, homology group, Rips complex, filtration, and persistence. In the second part, functions and models, which are based on the persistent homology results, are introduced to quantitatively analyze and compare the complicated topological information from different data.

\subsection{Persistent homology}
Persistent homology is a newly-invented model deeply rooted in algebraic topology, computational topology and combinatorial topology. In persistent homology, algebraic tools, such as quotient group, homology, exact sequence, etc, are used to characterize topological invariants, including connected components, circles, rings, channels, cavities, voids, etc. We present a very brief introduction in this section and refer interested readers to papers\cite{Edelsbrunner:2001,Edelsbrunner:2002,Zomorodian:2005} for more details.

\paragraph{Simplicial complex}
A $k$-simplex $\sigma^k=\{v_0,v_1,v_2,\cdots,v_k\}$ is the convex hull formed by $k+1$ affinely independent points $v_0,v_1,v_2,\cdots,v_k$ as follows,
\begin{eqnarray}\label{eq:couple_matrix1}
\sigma^k=\left\{\lambda_0 v_0+\lambda_1 v_1+ \cdots +\lambda_k v_k \mid \sum^{k}_{i=0}\lambda_i=1;0\leq \lambda_i \leq 1,i=0,1, \cdots,k \right\}.
\end{eqnarray}
Geometrically, a 0-simplex is a vertex, a 1-simplex is an edge, a 2-simplex is a triangle, and a 3-simplex represents a tetrahedron. 
Simplices are the building block for simplicial complex. In general, a simplicial complex $K$ is a finite set of simplices that satisfy two essential conditions. First, any face of a simplex from  $K$  is also in  $K$. Second, the intersection of any two simplices in  $K$ is either empty or shares faces. An oriented $k$-simplex $[\sigma^k]$ is a simplex together with an orientation, i.e., ordering of its vertex set. Graphs and networks, which are comprised of only vertices and edges, can be viewed as a simplicial complex with only 0-simplex and 1-simplex.

\paragraph{Homology}
A linear combination of $k$-simplexes forms a $k$-chain $c$, i.e., $c=\sum_{i}\alpha_i\sigma^k_i$. In computational topology, we usually assume $\{ \alpha_i \in Z_2 \} $. All $k$-chains from the simplicial complex $K$ together with addition operation (modulo-2) will form an Abelian group $C_k(K, \mathbb{Z}_2)$. Homomorphism can be defined between these Abelian groups. Particularly, a boundary operator $\partial_k: C_k \rightarrow C_{k-1}$ can be used to explore algebraic relations between two chain groups. More specifically, the boundary of an oriented $k$-simplex $[\sigma^k]=[v_0,v_1,v_2,\cdots,v_k]$ can be denoted as,
\begin{eqnarray}
\partial_k [\sigma^k] = \sum^{k}_{i=0} [ v_0, v_1, v_2, \cdots, \hat{v_i}, \cdots, v_k ].
\end{eqnarray}
Here $[v_0, v_1, v_2, \cdots ,\hat{v_i}, \cdots, v_k ]$ means a $(k-1)$ oriented simplex, which is generated by the original set of vertices $v_0,v_1,v_2,\cdots,v_{i-1},v_{i+1},\cdots, v_k$ except $v_i$. The boundary operator satisfies several properties, including $\partial_0= 0$ and $\partial_{k-1}\partial_k= 0$. With the boundary operation, we can define the $k$-th cycle group $Z_k$ and the $k$-th boundary group $B_k$ as follows,
\begin{eqnarray}
&& Z_k={\rm Ker}~ \partial_k=\{c\in C_k \mid \partial_k c=0\}, \\
&&  B_k={\rm Im} ~\partial_{k+1}= \{ c\in C_k \mid \exists d \in C_{k+1}: c=\partial_{k+1} d\}.
\end{eqnarray}
With $\partial_{k-1}\partial_k= 0$, cycle and boundary groups satisfy $B_k\subseteq Z_k$. And their quotient group $H_k=Z_k/B_k$ is known as the $k$-th homology group. The rank of $k$-th homology group $H_k$ is called $k$-th Betti number and satisfies,
\begin{eqnarray}
\beta_k = {\rm rank} ~H_k= {\rm rank }~ Z_k - {\rm rank}~ B_k.
\end{eqnarray}
Geometrically, we can regard $\beta_0$ as the number of isolated components, $\beta_1$ the number of one-dimensional loops, circles, or tunnels and $\beta_2$ the number of two-dimensional voids or holes. 

\paragraph{Rips complex and filtration}

\begin{figure}
\begin{center}
\begin{tabular}{c}
\includegraphics[width=0.8\textwidth]{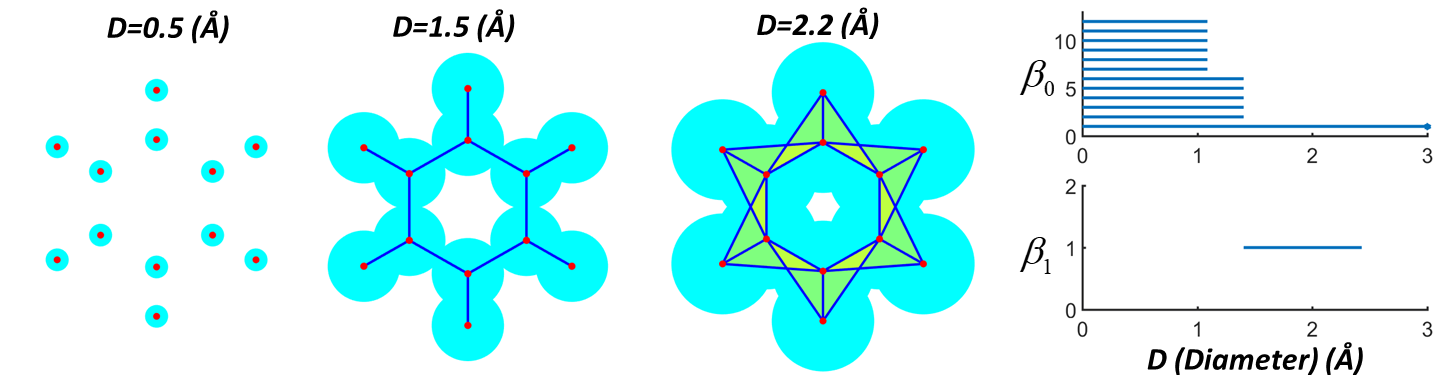}
\end{tabular}
\end{center}
\caption{ An illustration of persistent homology analysis of a benzene molecule. We consider carbon and hydrogen atoms in the benzene molecule as identical spheres (denoted as blue balls). The filtration parameter is chosen as the diameter (denoted as D) of these same-sized spheres. During a filtration process, a series of simplicial complexes are generated by systematically varying the diameter value. A simplical complex is made of simplexes. Geometrically, a 0-simplex is a vertex (red dot), a 1-simplex is an edge (blue line), a 2-simplex is a triangle (yellow triangle), a 3-simplex is a tetrahedron. The topological information of these simplicial complexes can be represented by barcodes. For the benzene example, we only demonstrate topological invariants $\beta_0$ and $\beta_1$. Invariant $\beta_0$ , representing the number of individual components, is plotted in the upper figure. Invariant $\beta_1$, denoting circle or ring structure, is shown in the lower figure. For both figures, the x-axis represents the filtration size, i.e., the diameter of the sphere, denoted as $D$ with unit \AA. The y-axis is the index of topological invariant. It can be seen that when $D=0.5$ \AA, all spheres are isolated from each other. Totally there are twelve individual components and zero rings or circles, therefore when $D=0.5$ \AA, $\beta_0=12$ and $\beta_1=0$. When diameter increase to $1.5$ \AA~ or $2.2$ \AA, all spheres are connected with each other and a ring is generated. Further enlargement of filtration size will result in more and more 2-simplex (triangles) and the ring will be blocked. In this way, when $D=3.0$ \AA, $\beta_0=1$ and $\beta_1=0$.
}
\label{fig:filtration}
\end{figure}
For any point set $X \in \mathbb{ R}^N$, one can associate each of its points with a closed ball (centered at the point) with diameter $\epsilon$. To explore the topological information embedded in the point set, Rips complex (or Vietoris-Rips complex) $\sigma$ can be generated. More specifically, a Rips simplex is formed when the largest distance between its any two vertices reaches $\epsilon$. Rips simplexes can then be combined together to generate Rips complex. It should be noticed that the generation of Rips complex requires a predefined value for diameter parameter $\epsilon$. However, how to find a suitable $\epsilon$ value which captures the underlying space of the point set is nontrivial. To solve this problem, the idea of filtration has been proposed\cite{Edelsbrunner:2002}. As illustrated in Figure \ref{fig:filtration},
instead of finding the best diameter value, an ever-increasing $\epsilon$ value is used to generate a series of topological spaces. These topological spaces can be represented by a nested sequence of simplicial complexes, and information of topological invariant can be derived from them. Each topological invariant will last for a certain $\epsilon$ values. The $\epsilon$ value, at which the invariant appears or disappears, is called birth time (BT) or death time (DT), respectively. In this way, each invariant has a ``lifespan", i.e., barcode length (BL), defined by its birth and death time. 

\paragraph{Persistent homology}
The nested sequence of complexes from the filtration process can be expressed as,
\begin{eqnarray}
\varnothing = K^0 \subseteq K^1 \subseteq \cdots \subseteq K^m=K.
\end{eqnarray}
And the $p$-persistent $k$-th homology group at filtration time $i$ can be represented as
\begin{eqnarray}
H^{i,p}_k=Z^i_k/(B_k^{i+p}\bigcap Z^i_k).
\end{eqnarray}
Essentially, persistence gives a geometric measurement of the topological invariant.

\subsection{Persistent homology based models and functions}
The pairs of BTs and DTs from the persistent homology analysis can be represented as barcodes in a graph. We can denote them as follows,
\begin{eqnarray}
\{ L_{k,j}=[a_{k,j}, b_{k,j}] | k=0,1,...; j=1,2,3,....,N_k \},
\end{eqnarray}
where parameter $k$ represent the $k$-dimension. For data points from Euclidean space, normally we only consider $k=0,1,2$. Parameter $j$ indicates the $j$-th topological invariant (barcodes) and $N_k$ is the number of $\beta_k$ topological invariant. The $a_{k,j}$ and $b_{k,j}$ are BTs and DTs. For simplification, we can define the set of $k$-th dimensional barcodes as,
$$ L_{k}= \{ L_{k,j}, j=1,2,3,....,N_k\}, \quad k=0, 1, ....$$

\paragraph{Persistent Betti functions}
Based on the persistent homology results, different functions are proposed to represent or analyze the topological information\cite{Edelsbrunner:2001,Carlsson:2009,bubenik:2015,Chintakunta:2015}. The persistent Betti number (PBN) is one of them, it is defined as the summation of all the $k$-th dimensional barcodes,
\begin{eqnarray}\label{eq:PBN}
f(x;L_{k})= \sum_{j} \chi_{[a_{k,j},b_{k,j}]}(x), \quad k=0, 1, ....
\end{eqnarray}
Function $\chi_{[a_{k,j}, b_{k,j}]}(x)$ is a step function, which equals to one in the region $[a_{k,j}, b_{k,j}]$ and zero otherwise. This equation transforms all the $k$-dimensional barcodes into a one-dimensional function. We can also define a scaled persistent Betti number (sPBN) as follows,
\begin{eqnarray}\label{eq:sPBN}
f(x;L_{k})= \frac{1}{N_k}\sum_{j} \chi_{[a_{k,j},b_{k,j}]}(x), \quad k=0, 1, ....
\end{eqnarray}

Another way of representing the barcode results is persistence landscapes\cite{bubenik:2015}. It is a function defined on each individual barcode $\{ L_{k,j}, k=0,1,...; j=1,2,3,....,N_k \}$,
\begin{eqnarray}\label{eq:couple_matrix25}
f(x,L_{k,j})=\begin{cases} \begin{array}{ll}
            0 &  {\rm if} \quad x \not\in (a_{k,j}, b_{k,j});\\
			x-a_{k,j}  & {\rm if} \quad x \in (a_{k,j}, (a_{k,j}+b_{k,j})/{2} ];\\		
            -x+b_{k,j} & {\rm if} \quad x \in [(a_{k,j}+b_{k,j})/{2}), b_{k,j}).
	      \end{array}
\end{cases}
\end{eqnarray}

However, these functions may not always be continuous, so another way of constructing persistent Betti function is proposed as follows,
\begin{eqnarray}\label{eq:PBF}
f(x;L_{k})= \sum_{j} e^{-\left(\frac{x- \frac{a_{k,j}+b_{k,j}}{2}}{w_j (b_{k,j}-a_{k,j})}\right)^2}, \quad k=0, 1, ....
\end{eqnarray}
Where $w_j$ are weight values, which can be adjusted based on different purposes\cite{Xia:2017similarity}.

To measure the disorder of a system, persistent entropy has been proposed\cite{Merelli:2015topological,Chintakunta:2015,Rucco:2016,Xia:2017multiscale}. For the $k$-dimensional barcodes, it is defined as,
\begin{eqnarray}\label{eq:filtrationM}
S_k=\sum_j^{N_k} - p_{k,j} ln(p_{k,j}), \quad k=0, 1, ...,
\end{eqnarray}
with the probability function
\begin{eqnarray}\label{eq:pi}
p_{k,j}=\frac{b_{k,j}-a_{k,j}}{\sum_j (b_{k,j}-a_{k,j})}, \quad k=0, 1, ....
\end{eqnarray}
The expression of persistent entropy can be simplified as follows,
\begin{eqnarray}\label{eq:MPE}
S_k=ln\left(\sum_j^{N_k} (b_{k,j}-a_{k,j})\right)- \frac{\sum_j^{N_k}\left( (b_{k,j}-a_{k,j}) ln(b_{k,j}-a_{k,j}) \right)}{\sum_j^{N_k} (b_{k,j}-a_{k,j})}, \quad k=0, 1, ....
\end{eqnarray}

\paragraph{Persistent similarity} The similarity between two barcodes can be measured by their distances\cite{Collins:2004,CEH07,cohen2010:lipschitz,bubenik:2015}.
The bottleneck distance between two sets of barcodes $L_{k_1}$ and $L_{k_2}$ is defined as
\begin{eqnarray}\label{eq:bottlenect}
d_B(L_{k_1},L_{k_2})=\underset{\gamma}{\inf} \quad \underset{j_1}{\sup} \| L_{k_1,j_1}-\gamma(L_{k_1,j_1}) \|_{\infty},
\end{eqnarray}
here $\gamma$ ranges over all bijections from barcodes $L_{k_1}$ to barcodes $L_{k_2}$. The distance between two barcodes $L_{k_1,j_1}$ and $L_{k_2,j_2}$ is defined as $\| L_{k_1,j_1}-L_{k_2,j_2} \|_{\infty}=\max \left\{ |a_{k_1,j_1}-a_{k_2,j_2}|, |b_{k_1,j_1}-b_{k_2,j_2}| \right\}$. Further the Wasserstein distance between two barcodes is defined as follows,
\begin{eqnarray}\label{eq:bottlenect}
d_W(L_{k_1},L_{k_2})=\underset{\gamma}{\inf} \left[ \sum_{j_1} \| L_{k_1,j_1}-\gamma(L_{k_1,j_1}) \|_{\infty}^p \right]^{\frac{1}{p}}.
\end{eqnarray}
Again $\gamma$ is bijection from barcodes $L_{k_1}$ to barcodes $L_{k_2}$. Parameter $p$ is a positive integer.

\section{Results and Discussions}\label{sec:results}

The major results from our persistent homology analysis of ion aggregates and hydrogen-bonding networks are presented in this section. In the first part, we illustrate the topological representation of data by two examples, i.e., NaCl crystal lattice structure and KSCN hydrogen-bonding network. We introduce the crystal lattice fingerprint (CLF) for the description of highly regular crystal lattice structures. CLF provides a unique topological representation of crystal lattices.

In the second part, we explore the ion aggregation mechanism by considering two solution systems, NaCl and KSCN solutions\cite{Kim:2014ionI,Choi:2014ionII}. By a systematical comparison of the results from two systems in various ion concentrations, we find that the proposed two ion aggregation types \cite{Kim:2014ionI} can be well characterized by our topological measurements. Essentially, the ion clusters can be characterized by a group of barcode located in the region from 3.0\AA~ to 4.0\AA. And the extended ion network is characterized by the assembly of the gigantic amount of barcodes in local regions.

In the last part, we study the hydrogen-bonding networks. Two different ways of choosing atoms are considered to build two types of networks, i.e., O-Network and ${\rm H_2O}$-Network. For both networks, we find that KSCN systems demonstrate a greater variation in their topological properties than NaCl. With concentration increase, large-sized local circles grow consistently in KSCN and the general distribution of circle sizes becomes more and more diverse. In contrast, a consistent decline of the average size of circle structures is observed in NaCl and the sizes of these circles become more and more uniformed. These topological features in ion aggregation and hydrogen-bonding networks have never been revealed before.

\subsection{Topological representation of data} \label{sec:top}

To have an intuitive understanding of topological invariants, we begin our introduction with topological fingerprints for highly regular crystal structures. Topological fingerprint is orginally proposed to characterize biomolecular structures\cite{KLXia:2014c,KLXia:2015a}. It provides a new structure representation that balances geometric details and topological simplicity. Topological fingerprint is highly efficient in describing consistent circle or loop structures within data. 

\paragraph{Topological fingerprint of crystal lattice structure}

\begin{figure}
\begin{center}
\begin{tabular}{c}
\includegraphics[width=0.8\textwidth]{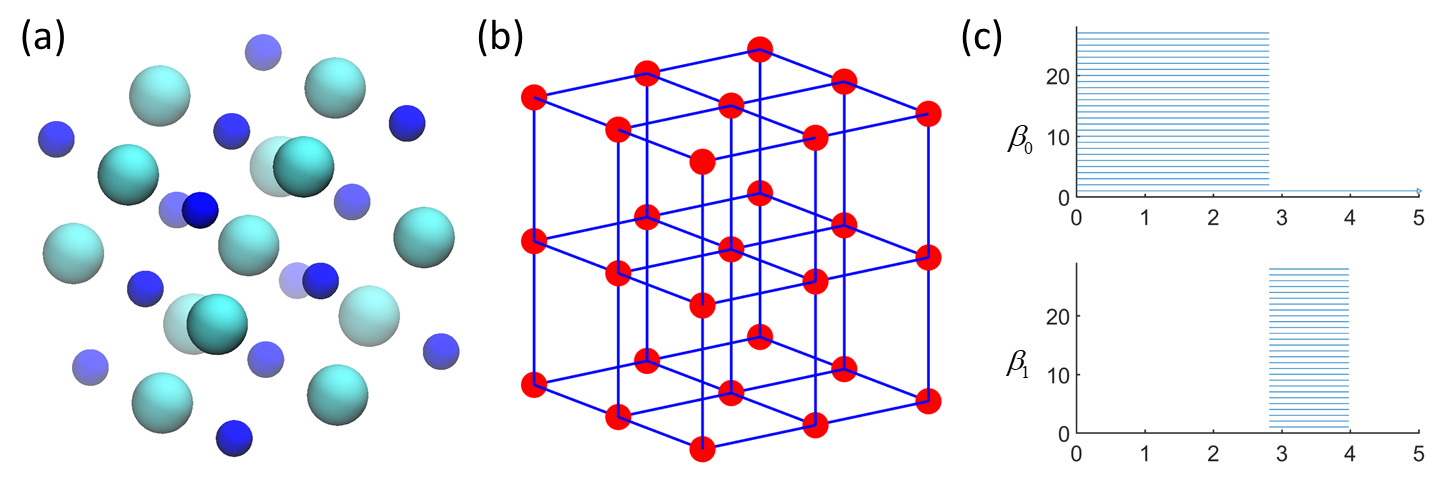}
\end{tabular}
\end{center}
\caption{Persistent homology analysis of crystal structure NaCl. ({\bf a}) The crystal structure of NaCl. In persistent homology analysis, all atoms, no matter Na or Cl, are treated equally as identical spheres with an ever-increasing diameter. Networks are constructed when these spheres are overlapped with each other. ({\bf b}) The network of NaCl crystal structure from persistent homology when the diameter of the sphere is 3.0\AA. ({\bf c}) Barcode representation for NaCl crystal structure. The number of $\beta_0$ bars is the number of atoms. There are totally 27 atoms, thus 27  $\beta_0$ bars. The length of $\beta_0$ bars usually represent bond length\cite{KLXia:2014persistent}. In this crystal structure, Na-Cl bond lengths are highly consistent with value around 2.80\AA~ as indicated by $\beta_0$ barlength. Further, $\beta_1$ bars are related the cycle, loop or ring structures in the multiscale networks built from the filtration process. As illustrated in ({\bf b}), the adjacent two Na atoms and two Cl atoms will form a rectangular ring, thus contribute to a $\beta_1$ bar. The length of $\beta_1$ bars is roughly the size of rings.
}
\label{fig:Crystal_lattice}
\end{figure}

A crystal structure is made of highly ordered particles. These particles aggregate together following certain symmetry rules, which play an import role in determining physical and chemical properties of the crystal structure. As persistent homology embeds geometric information into topological invariants, it can deliver a quantitative description of crystal lattice structures. We call this new representation as crystal lattice fingerprint (CLF).

An example of NaCl crystal structure is used to illustrate the crystal lattice fingerprint representation. A NaCl crystal with totally 27 atoms is shown in Fig \ref{fig:Crystal_lattice} ({\bf a}). In our persistent homology analysis, we do not discriminate different types of atoms, instead all atoms are treated equally as identical spheres with an ever-increasing diameter. Through a systematical variation of diameter value, a series of simplicial complexes are generated. Fig \ref{fig:Crystal_lattice} ({\bf b}) shows a simplicial complex produced at diameter 3\AA. This simplicial complex has only vertices and edges, thus equivalent to a network or graph. It can be seen that its structure is highly symmetric and all rectangular rings are of similar sizes. Our barcode results in Fig \ref{fig:Crystal_lattice} ({\bf c}) are consistent with this observation. In $\beta_0$ barcodes, there are totally 27 barcodes, representing 27 atoms in the system. All these bars are of equal size, about 2.80\AA, which  is bond length between atoms Na and Cl. It should be noticed that the length of $\beta_0$ bars means the size of individual components. Generally speaking, lengths of ``shorter" $\beta_0$ bars correspond to bond lengths, and lengths of ``longer" $\beta_0$ bars correspond to distances between ``clusters". In $\beta_1$ barcodes, all bars are identical with BT at 2.80\AA~ and DT at 3.95\AA. It should be noticed that $\beta_1$ BT is a diameter value, at which spheres overlap with adjacent ones to form a ring or circle. DT is when the ring or circle are ``blocked" by 2-simplexes (triangles). Stated differently, DT is diameter value at which the generated ring is filled up by triangles that are formed among adjacent three atoms. In our rectangular ring case, all BTs are equal to Na-Cl bond length about 2.80\AA~ and DTs are equal to the rectangular diagonal length, which is 3.95\AA. In this way, the $\beta_1$ BL provides a geometric measurement of ring or circle structures. It worth mentioning that the number of $\beta_1$ barcodes are not exactly the number of rings formed in the network. It can be seen that there are totally 36 rings in the network structure (as in Fig \ref{fig:Crystal_lattice} ({\bf b})), but only 28 $\beta_1$ bars. This is due to the reason that in algebraic representation, the missing 8 rings are merely linear combinations of the 28 intrinsic rings as represented in the barcode.

With its unique properties in structure characterization, the barcode representation works as the fingerprint of crystal lattice structure. It can been found that a unique CLF can be generated for each crystal lattice structure and two crystal lattice structures can never have the same CLF. This is due to the reason that our barcodes capture the intrinsic topological information of structures. The $\beta_0$ number gives a total number of atoms, $\beta_0$ BL gives bond length information. While $\beta_1$ characterizes how these atoms form ring and circle structures. It is worth mentioning that the graph or network model is just a simple case of simplicial complex. Essentially, simplicial complex is made from simplexes, such as vertices, edges, triangles, and tetrahedrons. If we remove all the higher dimensional simplexes like triangles and tetrahedrons, a simplicial complex will be reduced to a network (or graph). It can be observed more clearly when we consider irregular networks as the ones from ion aggregation and hydrogen bonding structures.

\paragraph{Topological representation of complicated networks}

\begin{figure}
\begin{center}
\begin{tabular}{c}
\includegraphics[width=0.8\textwidth]{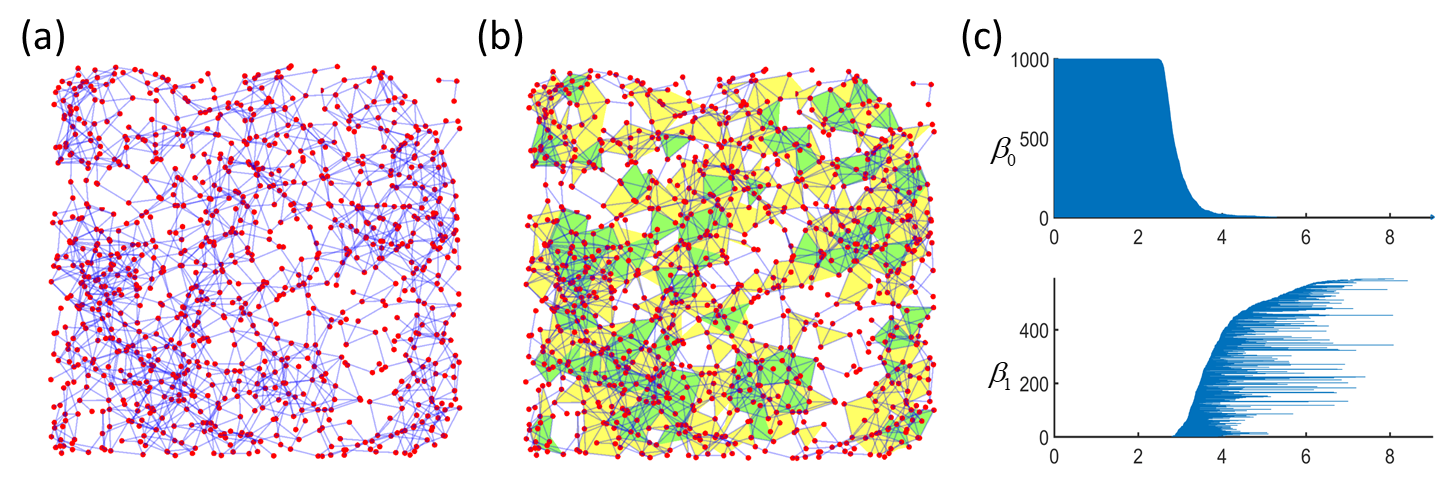}
\end{tabular}
\end{center}
\caption{Persistent homology analysis of hydrogen-bonding network. ({\bf a}) The illustration of hydrogen-bonding network from 10M KSCN solution. Only oxygen atoms are considered. The the hydrogen bond is specified if the distance between two oxygen atoms is larger than 2.5\AA~ and smaller than 3.5\AA\cite{Xenides:2006hydrogen,Marcus:2009effect}. ({\bf b}) The simplicial complex generated when filtration size is 3.5\AA. The network in ({\bf a}) is a special simplicial complex with only 0-simplex (vertices) and 1-simplex (edge). ({\bf c}) The barcode results of the hydrogen-bonding network from 10M KSCN solution. The barcode representation provides a new way to characterize the intrinsic topological information of hydrogen-bonding network, and it is free from hydrogen bond definition.
 }
\label{fig:Kscn_10M_O_example}
\end{figure}

Unlike the crystal lattice structure, ion aggregates and hydrogen-bonding networks are highly irregular and asymmetric. As discussed in the introduction, graph theory, particularly the spectral graph theory, is widely used to study these systems.  In this part, we will demonstrate that persistent homology provides a totally different perspective in characterizing highly complicated network structures.  Figure \ref{fig:Kscn_10M_O_example} ({\bf a}) demonstrates an example of hydrogen-bonding network. It is generated from 10 M KSCN solution and only oxygen atoms are considered. The detailed MD simulation setting and process is explained in Section \ref{sec:ion} and reference \cite{Kim:2014ionI,Choi:2014ionII}. The hydrogen bond is defined between two oxygen atoms if their distance is larger than 2.5\AA~ and smaller than 3.5\AA\cite{Xenides:2006hydrogen,Marcus:2009effect}. In contrast, a simplicial complex is shown in Fig \ref{fig:Kscn_10M_O_example} ({\bf b}). This simplicial complex is generated by setting the sphere diameter as 3.5\AA. Rough speaking, a simplicial complex is a network model combined with 2-simplexes (yellow triangles) and 3-simplexes (green tetrahedrons). More importantly, in persistent homology, a series of simplicial complexes can be systematically produced by varying a filtration parameter (i.e., sphere diameter). In this way, we do not need a predefined ``bond length" to build up a special network as in traditional models. The Betti number information can then be calculated from these simplicial complexes.

Barcode results for hydrogen-bonding network are shown in Figure \ref{fig:Kscn_10M_O_example} ({\bf c}). By comparing with barcodes from NaCl crystal lattice structure in Figure \ref{fig:Crystal_lattice} ({\bf c}),  it can be seen that distribution of bars in hydrogen networks are much more diverse and irregular. This indicates that both bond length and circle structure in a hydrogen-bonding network vary more dramatically. Further, it can be seen that some $\beta_0$ bars are much larger than $3.5$\AA. These bars no longer characterize bond length (as in definition, bond length should be smaller than 3.5\AA), instead they measure the relative distance between atom ``clusters".

\subsection{Ion aggregation} \label{sec:ion}

\begin{figure}
\begin{center}
\begin{tabular}{c}
\includegraphics[width=0.8\textwidth]{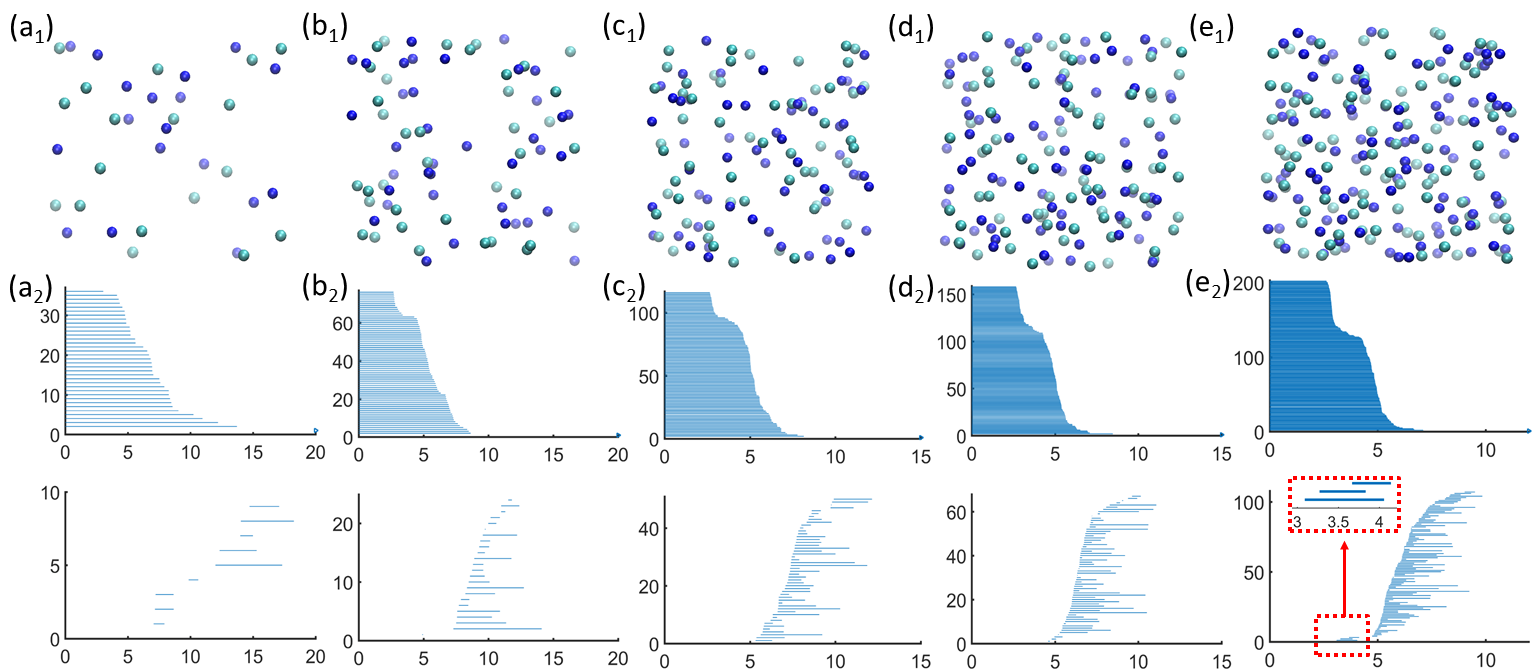}
\end{tabular}
\end{center}
\caption{The persistent barcodes for NaCl ion aggregation. (${\bf a}_1$) to (${\bf e}_1$) Configurations of NaCl ions at 1M, 2M, 3M, 4M and 5M concentration, respectively. (${\bf a}_2$) to (${\bf e}_2$) Persistent barcodes for NaCl ions at 1M, 2M, 3M, 4M and 5M concentration, respectively. It can be seen that total bar numbers, BTs, DTs, and BLs of $\beta_0$ and $\beta_1$ barcodes all change dramatically with ion concentration. At about 5M, NaCl ions reach saturation. At the same time, they form ion clusters, which is characterized by short bars around 3.0\AA~ to 4.0\AA. Moreover, there is a clear separation between these local bars and global bars. This unique barcode pattern can be viewed as a special fingerprint for ion aggregation through ion clusters. }
\label{fig:nacl_ions}
\end{figure}
To explore topological features in ion aggregation, we consider two well-studied systems, i.e., NaCl and KSCN, generated from the molecular dynamic simulation as test examples\cite{Kim:2014ionI,Choi:2014ionII}. The considered simulation trajectories are directly adopted from the references \cite{Kim:2014ionI,Choi:2014ionII} and general information of the data preparation are summarized as follows. For NaCl solutions, concentrations at 1.0M, 2.0M, 3.0M, 4.0M and 5.0M are considered. In MD simulations, each periodic box has 1000 TIP3P water molecules. And bar numbers of NaCl molecules at five different concentrations are 18, 38, 58, 79 and 101, respectively. For KSCN solutions, concentrations at 0.92M, 3.22M, 5.23M, 7.21M and 9.88M are studied. The same periodic box with 1000 TIP3P water molecules is used in MD. For five concentrations, bar numbers of KSCN molecules are 17, 70, 132, 215 and 395, respectively. For notational convenience, we round the KSCN concentration values to an integer and denote the five concentrations for KSCN as 1.0M, 3.0M , 5.0M, 7.0M and 10.0M, respectively. The force field parameters for ${\rm SCN}^-$ are obtained by using ANTECHARMBER module in AMBER 11 package. The atomic partial charges of ${\rm SCN}^-$ are obtained at the level of ${\rm HF/6-31G(d)}$ using Gaussian 03 package.  The atomic charges for S, C and N atoms are -0.560e, 0.145e and -0.585e, respectively. For non-bonding interactions, the cutoff distance is 10\AA. Long-range electrostatic interactions are treated by the particle-mesh Ewarld method in AMBER program. The energy minimization with steepest gradient descent method is employed in all KSCN and NaCl solution systems before MD simulation. Subsequently, the NPT ensemble with 1 atm pressure, 298K temperature and 1 fs time step are used. The simulation time is 2ns and the periodic box size is adjusted. Further, another 2 ns MD simulation with constant volume and temperature at 298K was performed for equilibration. Ions are allowed to diffuse in the periodic box. After that, a final 10ns simulation with NVT condition is employed. 
The highest concentrations for NaCl and KSCN are all close to their solubility limits at room temperature. We refer interested readers to references\cite{Kim:2014ionI,Choi:2014ionII} for more details.

We consider thirty configurations (or frames) obtained from each MD trajectory from the final equilibration process. We take ions from each configuration to form an ion aggregation system. O-Network and ${\rm H_2O}$-Network are constructed by selecting oxygen atoms or oxygen and hydrogen atoms from each configuration. Even though we call them O-Network and ${\rm H_2O}$-Network, they are just aggregations of atoms. There are no ``edges" or bonds in them, as we do not use a predefined bond length. In our persistent homology analysis, we do not distinguish between different ion types. Results for NaCl ions from the last frame of each MD trajectory are shown in Figure \ref{fig:nacl_ions}. Total bar numbers of $\beta_0$ barcodes are 36, 72, 116, 158 and 202, respectively. They are exactly twice the number of NaCl molecules. Circle information can be found in $\beta_1$ barcodes. We can see that the distribution of barcodes is consistently narrowing. In 1M concentration, largest $\beta_1$ DTs go as far as 18.0\AA.~ These largest DTs decrease to around 14.0\AA ~ in 2M solution results and further to around 12.5\AA, 11.0\AA~ and  9.5\AA ~ in 3M,  4M and 5M, respectively. More interestingly, in 5M concentration, short $\beta_1$ bars appear in the region between 3.0\AA~ to 4.0\AA~ as illustrated in the enlarged red box. And there is a clear separation between ``local" bars and ``global" bars. Simply speaking, local bars come earlier in the filtration process, whereas global bars appear much later. Local bars represent topological features at small local scale, while global bars are topological properties at large global scale. Take DNA molecule as an example, local $\beta_1$ bars represent circle structures from ribose sugar and base, while global $\beta_1$ bars represent circles formed between two layers of base pairs or minor and major grooves\cite{KLXia:2015d}. It has been suggested that NaCl and KSCN systems reflect two types of ion aggregations as their concentration approaches solubility limit\cite{Kim:2014ionI,Choi:2014ionII}. For NaCl ions, they aggregate by forming local ion clusters. From the above analysis,  we find that, topologically, local ion clusters are well characterized by our local $\beta_1$ bars. Moreover, a clear separation between local $\beta_1$ bars and global ones can be viewed as a topological fingerprint for this special type of ion aggregation.

\begin{figure}
\begin{center}
\begin{tabular}{c}
\includegraphics[width=0.8\textwidth]{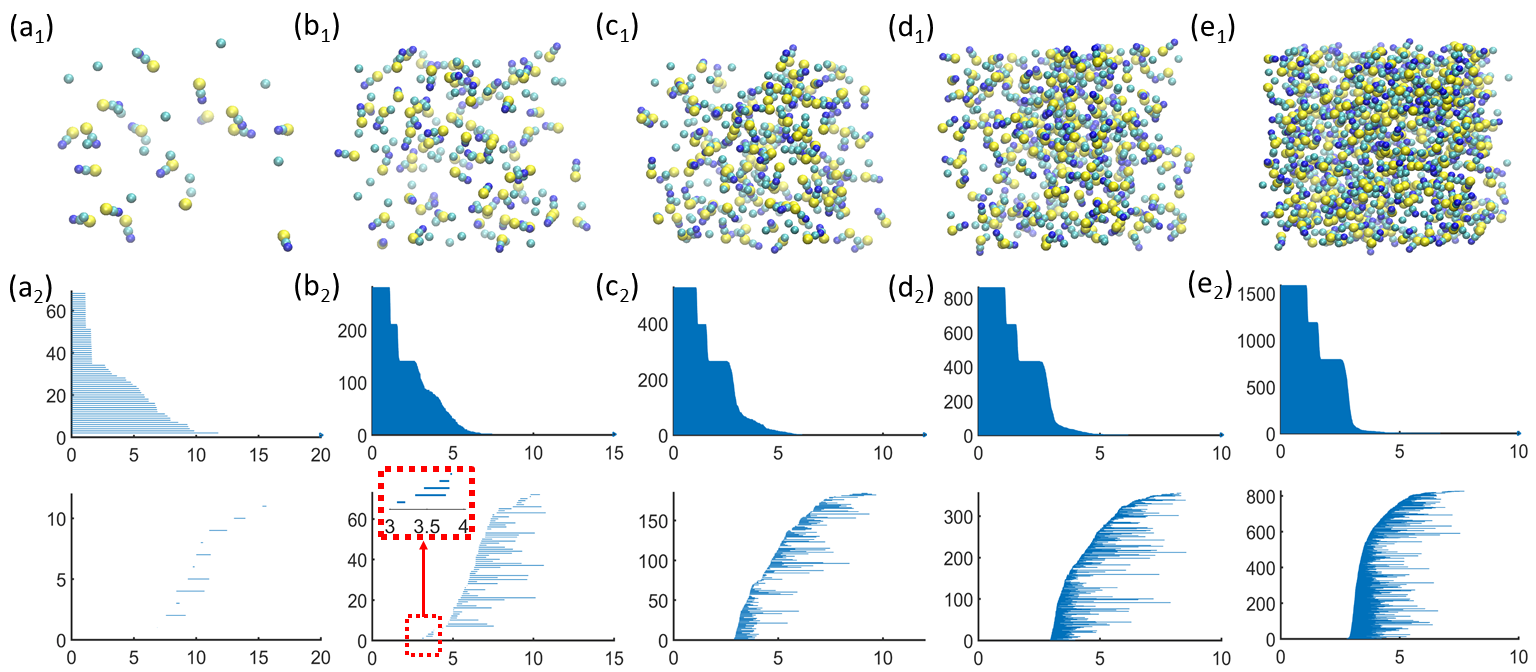}
\end{tabular}
\end{center}
\caption{The persistent barcodes for KSCN ion aggregation. (${\bf a}_1$) to (${\bf e}_1$) Configurations of KSCN ions at 1M, 3M, 5M, 7M and 10M concentration, respectively. (${\bf a}_2$) to (${\bf e}_2$) Persistent barcodes for NaCl ions at 1M, 3M, 5M, 7M and 10M concentration, respectively. It can be seen that total bar numbers, BTs, DTs, and BLs  of $\beta_0$ and $\beta_1$ all change dramatically with the ion concentration. KSCN solution reaches solubility limit at around 10M, when KSCN ions form ``extended ion network". Topologically, ``extended ion network" is a highly compact configuration, which is characterized by the accumulation of only local circle structures. Further, ion clusters, indicated by short barcodes, also appear in 3M KSCN concentration.
}
\label{fig:kscn_ions}
\end{figure}

In contrast, barcode results for KSCN ions from the last frame of each MD trajectory are shown in Figure \ref{fig:kscn_ions}. As expected, $\beta_0$ bar numbers equal to total numbers of ions, which are precisely four times the KSCN molecule numbers. For $\beta_1$ barcodes, a shrinking of barcode distributions with ion concentrations, as seen in NaCl systems, is also observed.  Further, Figs \ref{fig:kscn_ions}(${\bf c}_2$) to (${\bf e}_2$) demonstrate a clear pattern that with the increase of concentration, more and more bars emerge in ``local" regions with filtration value from 3.0\AA~ to 4.0\AA. Especially when ion concentration is 10M, ion networks are dominated by these local circle structures, with almost no significant global features. This is dramatically different from NaCl ion aggregation as in Fig \ref{fig:nacl_ions}(${\bf e}_2$). This type of ion aggregation is called ``extended ion network" \cite{Kim:2014ionI,Choi:2014ionII}. From our barcode results, it can be seen that ions form a highly compact configuration with a gigantic amount of local circles but no global circles. Interestingly, the barcode pattern from 3M KSCN concentration is very similar to that from 5M NaCl concentration. KSCN ions also form local clusters as indicated by the short barcodes in Figure \ref{fig:kscn_ions}(${\bf b}_2$). However, unlike NaCl systems, which reach saturation at 5M, KSCN ions have a much higher solubility limit and demonstrate a unique aggregation pattern.

From the above discussions, we can find that the difference in NaCl and KSCN ion aggregation is quite dramatic in terms of their topological measurements, including $\beta_0$ and $\beta_1$ bar numbers, BTs, DTs, and BLs. We have analyzed all the other 29 frames, similar patterns have been observed. To further explore the ion aggregation, particular the mutual influence between hydrogen-bonding networks and ion specificity, we consider topological properties in O-Networks and ${\rm H_2O}$-Networks. The results are presented in the following sections.

\subsection{Hydrogen-bonding network} \label{sec:hydrogen}

\begin{figure}
\begin{center}
\begin{tabular}{c}
\includegraphics[width=0.8\textwidth]{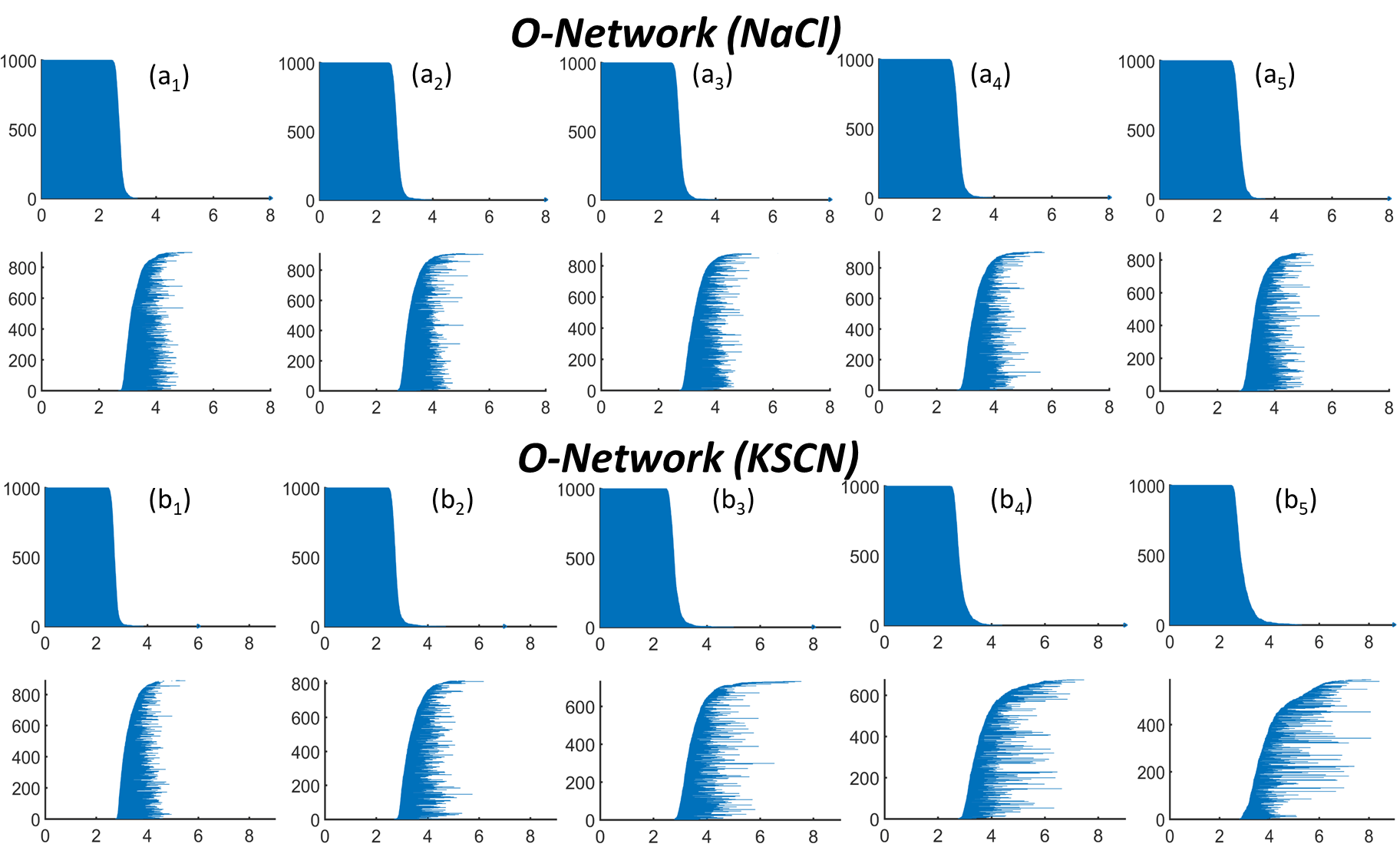}
\end{tabular}
\end{center}
\caption{The persistent barcodes for NaCl and KSCN oxygen-networks (O-networks). (${\bf a}_1$) to (${\bf a}_5$) Persistent barcodes for NaCl O-networks at ion concentrations 1M, 2M, 3M, 4M and 5M, respectively. (${\bf b}_1$) to (${\bf b}_5$) Persistent barcodes for KSCN O-networks at ion concentrations 1M, 3M, 5M, 7M and 10M, respectively. It can be seen clearly that, total bar number of $\beta_1$ barcodes for NaCl O-networks remains relatively stable at different ion concentrations. In contrast, $\beta_1$ barcodes for KSCN O-networks vary dramatically at different ion concentrations.
}
\label{fig:oxgen_barcodes}
\end{figure}

\begin{figure}
\begin{center}
\begin{tabular}{c}
\includegraphics[width=0.5\textwidth]{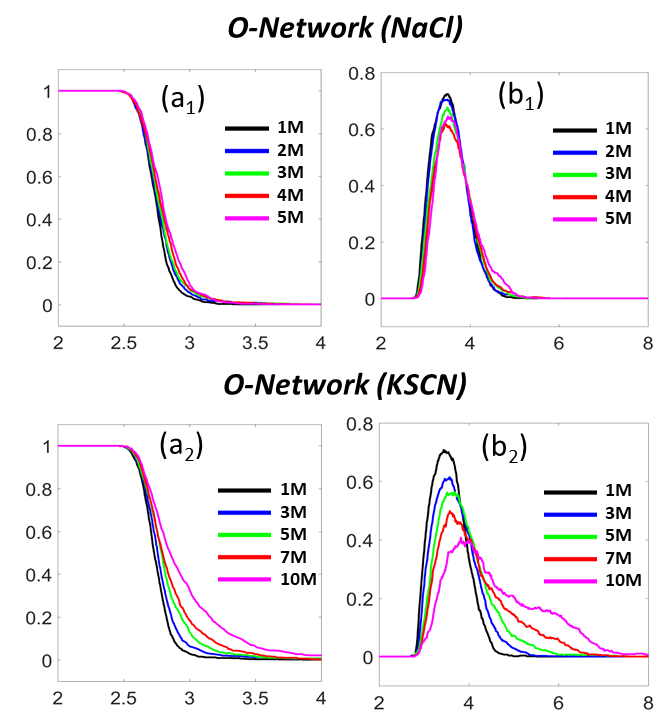}
\end{tabular}
\end{center}
\caption{The illustration of scaled persistent Betti number (sPBN) for NaCl and KSCN O-Networks. Generally speaking, as ion concentration increases, values of $\beta_0$ sPBN increase consistently. While values of $\beta_1$ sPBN decrease at earlier times of filtration, then increase at later times of filtration. Values of KSCN sPBNs change much more dramatically than NaCl ones.
}
\label{fig:pbn_O}
\end{figure}

\begin{figure}
\begin{center}
\begin{tabular}{c}
\includegraphics[width=0.8\textwidth]{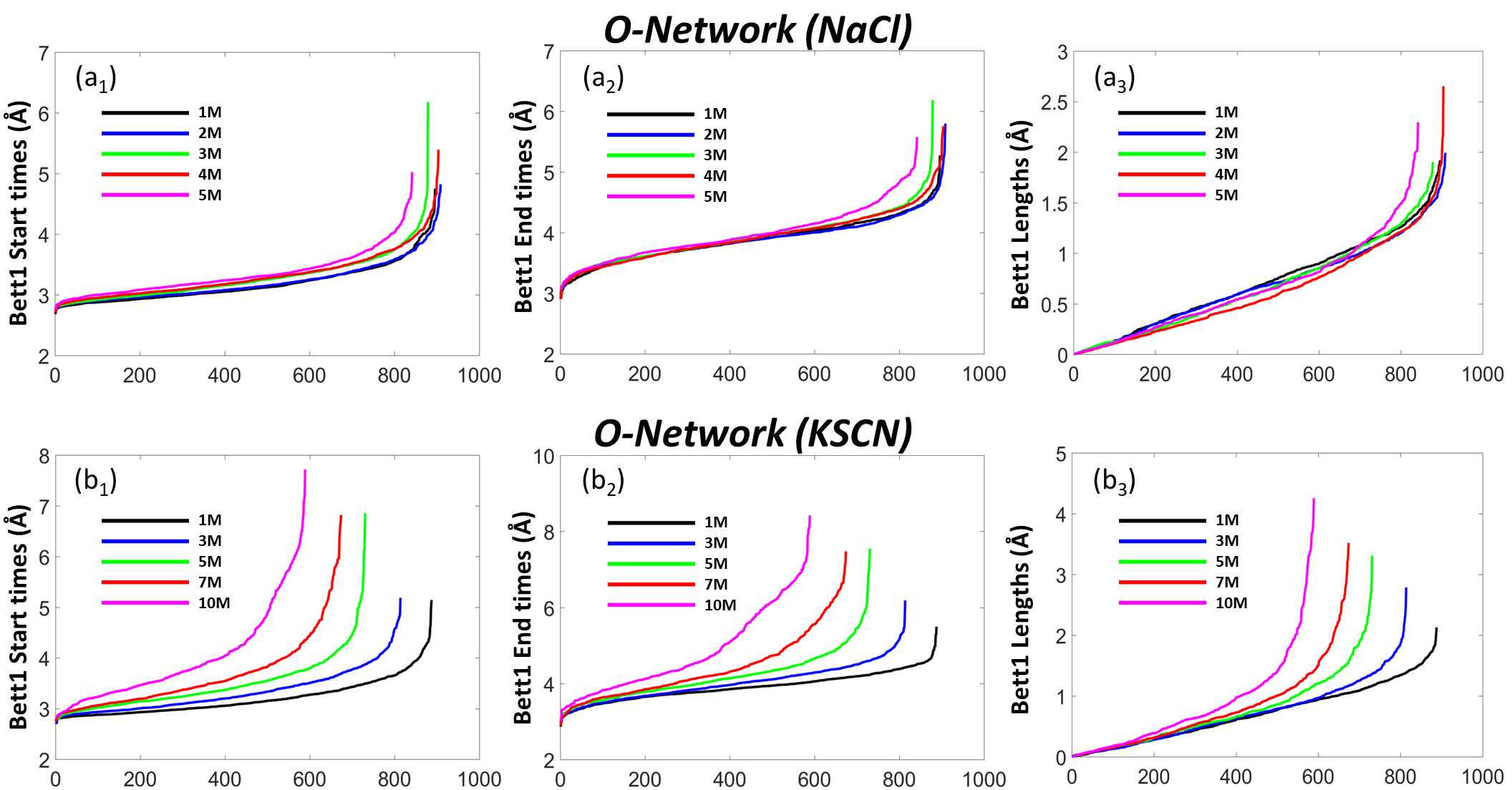}
\end{tabular}
\end{center}
\caption{Three persistent barcode measurements, i.e., BT, DT, and BL, for NaCl and KSCN O-networks. For NaCl, results for BTs, DTs, and BLs are shown in (${\bf a}_1$) to (${\bf a}_3$), respectively. For KSCN, results for three measurements are illustrated in (${\bf b}_1$) to (${\bf b}_3$), respectively. All measurement values are rearranged in an ascending order. The x-axis is new barcode index after rearrangement. For NaCl systems, at all five different concentrations, the variation of their persistent measurements are relatively small. The total numbers of $\beta_1$ bars are 896, 909, 879, 904 and 842, respectively. The distributions of BTs, DTs, and BLs, are rather consistent with less variation. In contrast, the measurements of KSCN systems differ greatly in different concentrations. The total bar number decreases dramatically from 888, 814, 731, 674 to 589. More interestingly, the average size of ring or circle structures in the systems increase with ion concentration.
}
\label{fig:O_sel}
\end{figure}

\begin{table}[htbp]
  \centering
	\caption{The comparison of topological measurements for NaCl and KSCN O-Networks from different concentrations. Thirty configurations from the equilibrium state are used in persistent homology analysis. The parameter ``C" denotes the concentration, its unit is mole. Parameter $N_{\beta_1}$ represents the average value of total number of $\beta_1$ bars. Notations ``BT", ``DT" and ``BL" represent birth time, death time and bar length, respectively. The unit for them is \AA. Notations ``M" and ``V" represent mean value and variance value. For instance, ``M(BT)" is mean value of birth times (over all barcodes from thirty configurations). For NaCl systems, with concentration rise, total number of $\beta_1$ bars slightly and consistently decreases from 904 to 849. Even though mean value of $\beta_1$ BLs decrease steadily, variance of $\beta_1$ BLs remains relatively stable at around 0.195\AA. In contrast, for KSCN systems, their total number of $\beta_1$ bars decreases quickly from 893 to 603, whereas mean and variance of $\beta_1$ BLs increase. Particularly, variance value has a huge rise. Topologically, this means that, with concentration increase, number of circles or loops consistently reduces in both NaCl and KSCN O-Networks. But KSCN O-Networks decrease much faster. Moreover, for NaCl systems, even thought the average size of circle structures consistently decrease, the variation of their sizes remains considerable stable. On the contrary, for KSCN systems, the increase of mean and variance of the circle sizes indicates the generation of more and more large-sized circles with ion concentration. }
    \begin{tabular}{|c|c|c|c|c|c|c|c|} \hline
 \multicolumn{8}{|c|}{NaCl O-Network} \\ \hline
  C & $N_{\beta_1}$ & M(BT)  &V(BT) & M(DT) &V(DT) & M(BL) &V(BL)\\ \hline
 1 &904 &3.192 &0.111 &3.891 &0.129 &0.699 &0.194\\ \hline
 2 &894 &3.233 &0.117 &3.899 &0.138 &0.667 &0.192\\ \hline
 3 &880 &3.263 &0.118 &3.912 &0.146 &0.649 &0.192\\ \hline
 4 &868 &3.301 &0.127 &3.935 &0.163 &0.633 &0.195\\ \hline
 5 &849 &3.340 &0.140 &3.969 &0.187 &0.629 &0.204\\ \hline
  \multicolumn{8}{|c|}{KSCN O-Network} \\ \hline
  C & $N_{\beta_1}$ & M(BT)  &V(BT) & M(DT) &V(DT) & M(BL) &V(BL) \\ \hline
  1 &893 &3.196 &0.123 &3.906 &0.137 &0.710 &0.201 \\ \hline
 3 &816 &3.327 &0.180 &4.026 &0.222 &0.699 &0.242 \\ \hline
 5 &751 &3.445 &0.264 &4.160 &0.367 &0.715 &0.303 \\ \hline
 7 &691 &3.612 &0.418 &4.351 &0.621 &0.739 &0.391 \\ \hline
10 &604 &3.969 &0.824 &4.785 &1.201 &0.816 &0.559 \\ \hline
    \end{tabular}
  \label{tab:o-networks}
\end{table}

We consider topological features of two sets of points. The first one is made of only oxygen atoms, we call it O-Network. As mentioned above, traditional graph and network models employ a special hydrogen bond length to build up a network among oxygen atoms. With filtration process, persistent homology analysis is free from this predefined bond length. The other one comprises both oxygen and hydrogen atoms, it is denoted as ${\rm H_2O}$-Networks.

Unlike the ion aggregation systems, barcode results for hydrogen-bonding networks remain relatively stable with concentration variations. However, the difference between two types of ion aggregation and between ions in different concentrations can still be discerned and characterized by our topological models. To facilitate presentation, we will show barcode results for the last frame of each MD trajectory first, and later summarize results from all thirty configurations in Tables.

\paragraph{Networks with only oxygen atoms}
Barcode results for NaCl and KSCN O-Networks from the last frame are shown in Fig \ref{fig:oxgen_barcodes}. Figs \ref{fig:oxgen_barcodes} (${\bf a}_1$) to (${\bf a}_5$) are for NaCl systems with ion concentrations from 1M to 5M, respectively. And Figs \ref{fig:oxgen_barcodes} (${\bf b}_1$) to (${\bf b}_5$) are for KSCN systems with concentration 1M, 3M, 5M, 7M and 10M, respectively. It can be seen that NaCl and KSCN $\beta_0$ barcodes are rather similar. They all have totally 1000 bars and their DTs are all distributed in the region from 2.5\AA~ to 3.5\AA. In contrast, NaCl and KSCN $\beta_1$ barcodes differ greatly. For all NaCl systems, the total numbers of $\beta_1$ bars are relatively consistent with values around 800 to 900. However, KSCN $\beta_1$ bar numbers vary dramatically from 600 to 900. To quantitatively compare these barcodes, scaled persistent Betti number as in Eq (\ref{eq:sPBN}) is considered and illustrated in Fig \ref{fig:pbn_O}. Roughly speaking, for both NaCl and KSCN systems, with ion concentration increase, values of $\beta_0$ sPBNs increase consistently. Values for $\beta_1$ functions decrease when filtration size is from about 3.0\AA~ to 4.0\AA, then keep increasing when size is larger than 4.0\AA. The variation of KSCN sPBNs with ion concentration is much larger than that of NaCl.

With the great importance of $\beta_1$ bars in circle structure characterization, we further consider distributions of $\beta_1$ topological measurements, including BT, DT, and BL. For each measurement, we arrange its values in an ascending order. Results are plotted in Fig \ref{fig:O_sel}. It can be seen that for NaCl systems distributions of $\beta_1$ BLs are very stable in various concentrations, with only small fluctuations in largest $\beta_1$ BLs. This indicates that sizes of local circle structures are very stable with no dramatic change with concentration.  On the contrary, distributions of $\beta_1$ BLs for KSCN changes consistently. With concentration rise, the proportion of large BLs in $\beta_1$ barcodes increases steadily. Sizes of large $\beta_1$ BLs also increase consistently. This means that KSCN O-Networks undergo a dramatic change with more and more large-sized local circles being consistently generated. In general, KSCN has a much greater variation with ion concentration in all three topological measurements.

We summarize in Table \ref{tab:o-networks} the total number of $\beta_1$ bars, mean and variance of BTs, DTs, and BLs from all thirty configurations. It can be seen that, with concentration rise, mean and variance of BTs and DTs for both NaCl and KSCN consistently increase. This means a systematical shift of $\beta_1$ bars to the later time of filtration. Difference between NaCl and KSCN ion aggregation can be described quantitatively by their $\beta_1$ bar numbers and bar sizes. More specifically, for KSCN systems, with concentration rise, their $\beta_1$ bar number decreases dramatically, whereas both mean and variance of $\beta_1$ BLs increase. For NaCl systems, their $\beta_1$ bar number shows a moderate decrease with ion concentration. Further, their mean value consistently decreases, but variance remains relatively stable. Topologically, it means that, with the increase of ion concentration, both NaCl and KSCN O-Networks suffer a loss of local circle structures, whereas KSCN O-Networks are more severe. Moreover, in NaCl systems, sizes of circles become smaller and smaller with concentration rise, while for each individual concentration all circles have a similar size. On the contrary, in KSCN systems, not only sizes of these local circle structures become larger and larger with concentration, but also the difference among these local circle structures at each individual concentration.

It has been suggested that water hydrogen-bonding structures in high concentration solutions are morphologically similar to ion network structures in high KSCN solutions. So we calculate the mean and variance of BTs, DTs, and BLs of KSCN ion networks generated from thirty configurations from 10M concentration. The corresponding values are 3.619, 0.589, 4.367, 0.771, 0.748 and 0.421. The total number of $\beta_1$ bars are 813. It can be seen that they do share a great similarity with topological measurements from KSCN O-Networks at 10M.

\begin{figure}
\begin{center}
\begin{tabular}{c}
\includegraphics[width=0.8\textwidth]{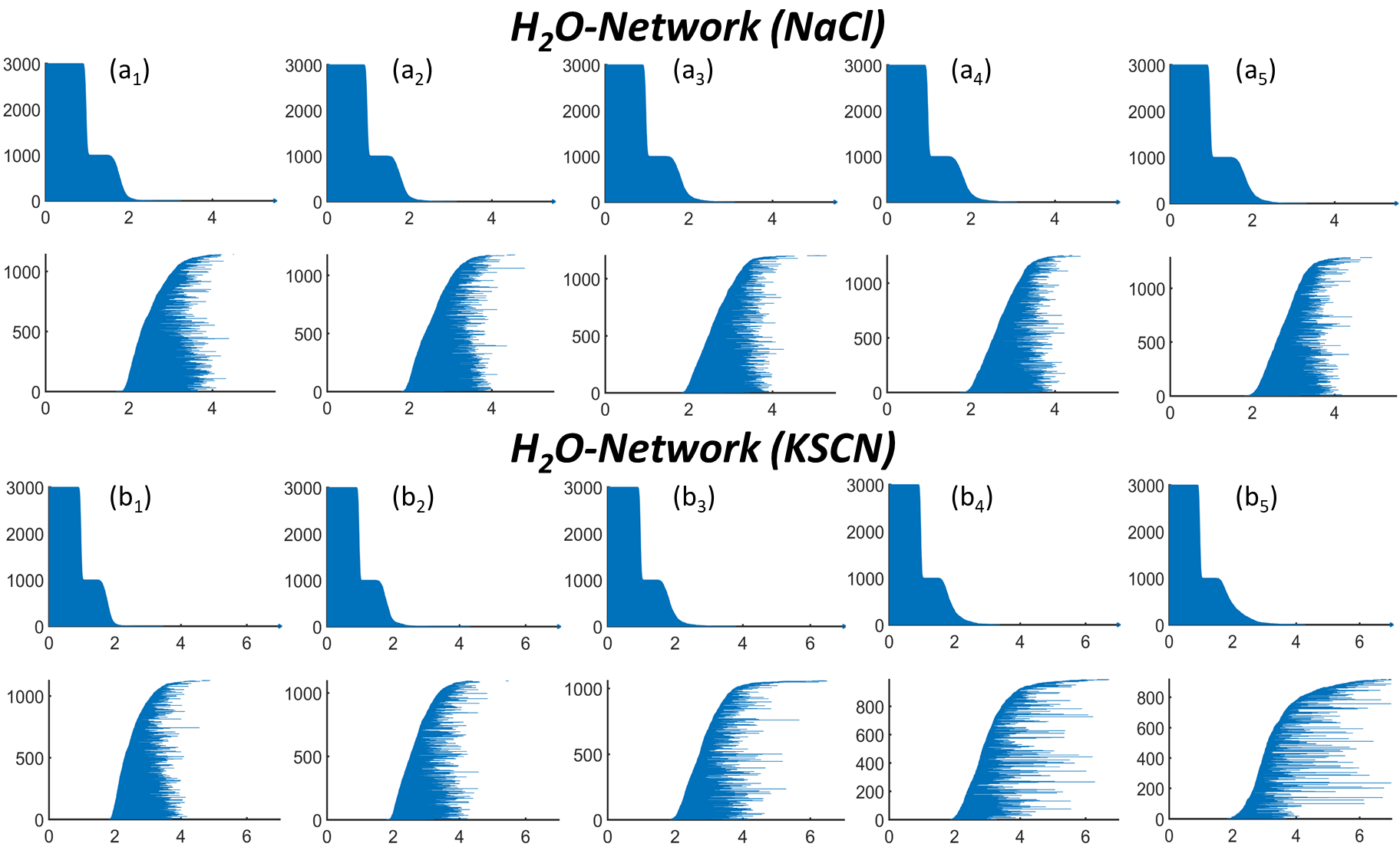}
\end{tabular}
\end{center}
\caption{The persistent barcodes for NaCl and KSCN ${\rm H_2O}$-networks. (${\bf a}_1$) to (${\bf a}_5$) Persistent barcodes for NaCl ${\rm H_2O}$-networks at ion concentrations 1M, 2M, 3M, 4M and 5M, respectively. (${\bf b}_1$) to (${\bf b}_5$) Persistent barcodes for KSCN ${\rm H_2O}$-networks at ion concentrations 1M, 3M, 5M, 7M and 10M, respectively. It can be seen clearly that, the total number of $\beta_1$ barcodes for NaCl ${\rm H_2O}$-networks remains relatively stable at different ion concentrations. In contrast, $\beta_1$ barcodes for KSCN ${\rm H_2O}$-networks vary dramatically at different ion concentrations.
}
\label{fig:water_barcodes}
\end{figure}

\begin{figure}
\begin{center}
\begin{tabular}{c}
\includegraphics[width=0.5\textwidth]{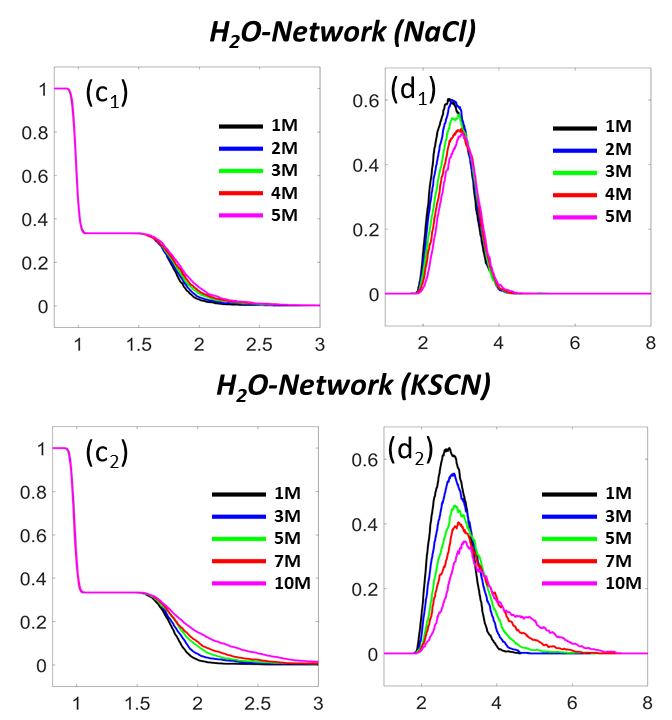}
\end{tabular}
\end{center}
\caption{The illustration of scaled persistent Betti number (sPBN) for NaCl and KSCN ${\rm H_2O}$-Networks. Generally speaking, with the increase of ion concentration, the values of $\beta_0$ sPBN increase consistently. While values of $\beta_1$ sPBN decrease at earlier times of filtration, then increase at later times of filtration. Values of KSCN sPBNs change much more dramatically than NaCl ones.
}
\label{fig:pbn_H2O}
\end{figure}

\begin{figure}
\begin{center}
\begin{tabular}{c}
\includegraphics[width=0.8\textwidth]{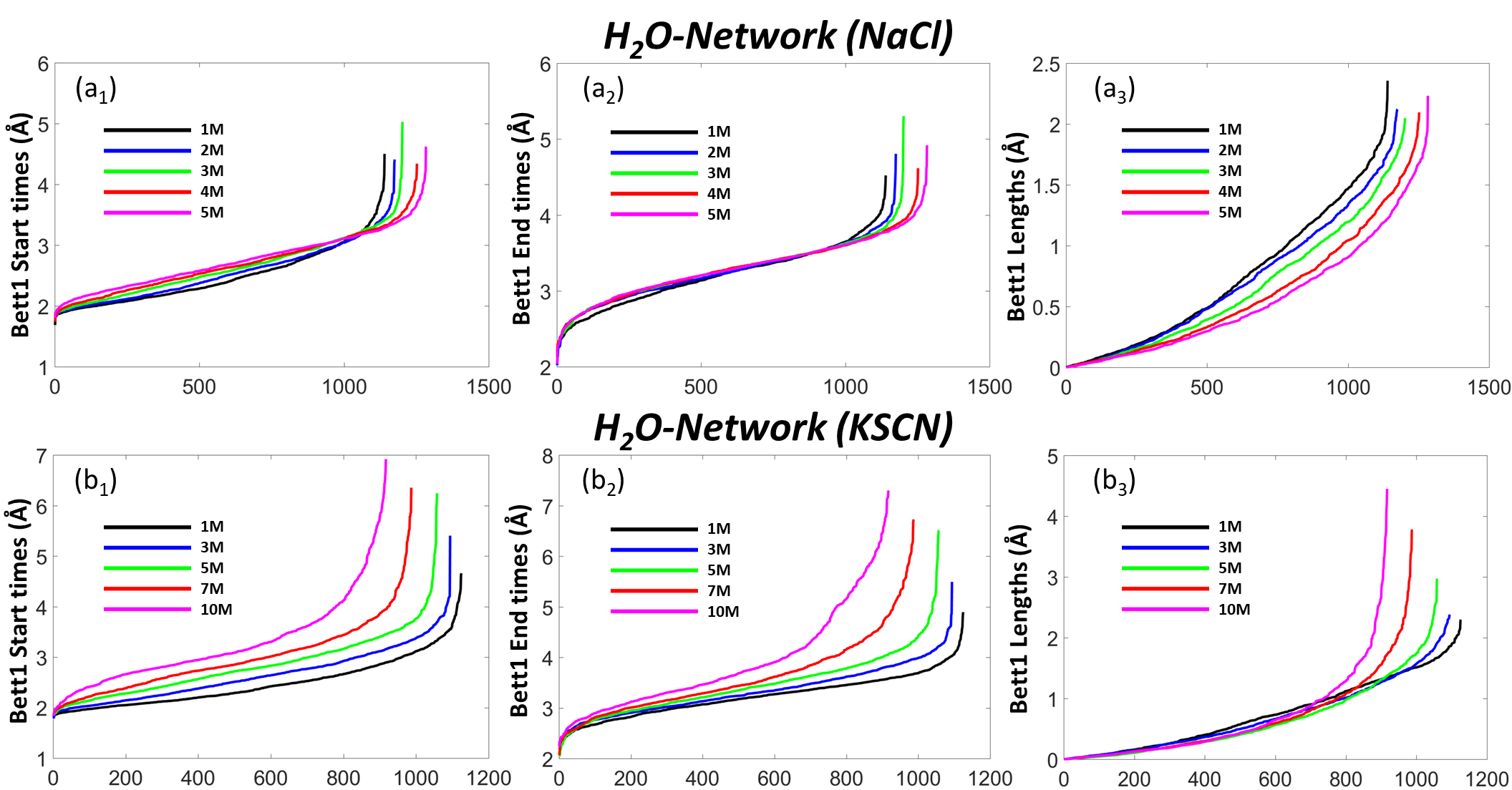}
\end{tabular}
\end{center}
\caption{ Three persistent barcodes measurements, i.e., BT, DT, and BL, for NaCl and KSCN ${\rm H_2O}$-Networks. For NaCl, the results for BTs, DTs, and BLs, are shown in ( ${\bf a}_1$) to (${\bf a}_3$), respectively. KSCN results are shown in ( ${\bf b}_1$) to (${\bf b}_3$), respectively. Measurement values are rearranged in an ascending order. The x-axis is new barcode index after rearrangement. Unlike NaCl O-Network results as illustrated in Fig \ref{fig:O_sel}, a much larger variation of $\beta_1$ bars are observed in NaCl ${\rm H_2O}$-Networks systems. The total numbers of bars for five different concentrations are 1140, 1175, 1202, 1252 and 1283, respectively, indicating a clear increase with ion concentration. However, distributions of BTs and DTs have less variation, which is consistent with NaCl O-Networks.
Further, the behaviors of KSCN ${\rm H_2O}$-Networks systems are similar to KSCN O-Networks. The total bar numbers decrease dramatically with concentration. Their values are 1125, 1094, 1058, 987 and 917 respectively. The average size of ring or circle structures in these systems increase with ion concentration.
}
\label{fig:water_sel}
\end{figure}

\begin{table}[htbp]
  \centering
	\caption{The comparison of persistent measurements for NaCl and KSCN ${\rm H_2O}$-Networks in different concentrations. Parameter ``C" denotes the concentration, its unit is mole. Parameter $N_{\beta_1}$ represents total number of $\beta_1$ bars. Notations ``BT", ``DT" and ``BL" represent birth time, death time and bar length, respectively. The unit for them is \AA. Notation ``M" and ``V" means the mean value and variance value, respectively. For instance, ``M(BT)" is mean value of birth time. For KSCN systems, as ion concentration increases, their $\beta_1$ bars decrease in numbers and average sizes, while increase in variance of sizes.
For NaCl systems, total number of $\beta_1$ bars rigorously increase with ion concentration, while mean and variance of the bar sizes constantly decrease. Topologically, it means that, with ion concentration increase, circle structures in KSCN ${\rm H_2O}$-Networks decrease greatly in their bar number and average size. More importantly, these circles becomes more and more diverse in their sizes. In contrast, circle structures in NaCl ${\rm H_2O}$-Networks consistently increase with concentration. More interestingly, their sizes not only become smaller, but also more uniformed.}
\begin{tabular}{|c|c|c|c|c|c|c|c|} \hline
 \multicolumn{8}{|c|}{NaCl ${\rm H_2O}$-Network} \\ \hline
  C & $N_{\beta_1}$ & M(BT)  &V(BT) & M(DT) &V(DT) & M(BL) &V(BL) \\ \hline
 1 &1145 &2.489 &0.216 &3.226 &0.148 &0.737 &0.304 \\ \hline
 2 &1182 &2.562 &0.213 &3.242 &0.146 &0.680 &0.281  \\ \hline
 3 &1215 &2.629 &0.210 &3.266 &0.144 &0.636 &0.263  \\ \hline
 4 &1238 &2.685 &0.199 &3.283 &0.144 &0.598 &0.243  \\ \hline
 5 &1247 &2.732 &0.192 &3.306 &0.151 &0.574 &0.233  \\ \hline
 \multicolumn{8}{c|}{KSCN ${\rm H_2O}$-Network} \\ \hline
  C & $N_{\beta_1}$ & M(BT)  &V(BT) & M(DT) &V(DT)  & M(BL) &V(BL)\\ \hline
 1 &1112 &2.474 &0.229 &3.236 &0.165 &0.761 &0.315  \\ \hline
 3 &1094 &2.658 &0.273 &3.341 &0.226 &0.683 &0.317  \\ \hline
 5 &1043 &2.786 &0.324 &3.440 &0.326 &0.654 &0.331  \\ \hline
 7 &1000 &2.944 &0.417 &3.573 &0.505 &0.629 &0.363  \\ \hline
10 &899 &3.233 &0.688 &3.876 &0.967 &0.643 &0.460  \\ \hline
    \end{tabular}
  \label{tab:h2o-networks}
\end{table}

\paragraph{Networks with oxygen and hydrogen atoms}
To further explore topological properties of the NaCl and KSCN systems, we consider the NaCl and KSCN ${\rm H_2O}$-Networks. Results from the last frame of each MD trajectory are shown in Fig \ref{fig:water_barcodes} with same notations as O-Networks. As expected, patterns of $\beta_0$ barcodes are much more stable than $\beta_1$ barcodes. For all systems, there are always 3000 $\beta_0$ bars, indicating 3000 atoms. Among these bars, 2000 bars have BLs around 1.0\AA, representing all H-O bonds within ${\rm H_2O}$ molecule. The other 1000 bars have BLs around 1.50\AA~ to 3.00\AA. They indicate the relative distances between adjacent ${\rm H_2O}$ molecules. Further, total bar number of NaCl $\beta_1$ barcodes consistently increases with concentration, whereas KSCN $\beta_1$ barcode number declines steadily. To facilitate a better comparison, we consider scaled persistent Betti numbers and illustrate the results in Fig \ref{fig:pbn_O}. Again, for both NaCl and KSCN systems, values of $\beta_0$ sPBNs increase with ion concentration. Their $\beta_1$ sPBNs decrease when filtration size is around 2.0\AA~ to 3.5\AA, then increase when filtration size is larger than 3.5\AA. KSCN sPBNs show a much larger variation than NaCl sPBNs.

Similar to the analysis in O-Networks, we plot the three topological measurements of $\beta_1$ barcodes in an ascending order in Fig \ref{fig:water_sel}. As expected, KSCN ${\rm H_2O}$-Networks shows a larger variation with ion concentration in all three measurements. 
However, different from the situations in O-Networks. A clear increase of barcodes with ion concentration is observed in NaCl ${\rm H_2O}$-Networks. Moreover, for KSCN ${\rm H_2O}$-Networks, even though there is a consistent increase of $\beta_1$ bars with larger BLs, there is also a simultaneous increase of $\beta_1$ bars with smaller BLs. This means the size gap between BLs gradually deepens, as concentration increase.

To facilitate a more quantitative comparison, statistic properties of $\beta_1$ BTs, DTs, and BLs from all thirty configurations are summarized in Table \ref{tab:h2o-networks}. It can be seen that, as ion concentration increases, mean values of BTs and DTs for both NaCl and KSCN increase steadily. This is consistent with results from NaCl and KSCN O-Networks. However, even though both BT and DT variances for KSCN show a uniform rise, BT and DT variance for NaCl either declines or fluctuates around 0.15\AA. Further, for NaCl systems, their $\beta_1$ number increase with ion concentration, whereas the mean and variance of BL decrease. For KSCN systems, as concentration rises, their $\beta_1$ number and the average size of BL decreases consistently, while the variance of BL increases. General speaking, the above results indicate that with concentration increase, KSCN ${\rm H_2O}$-Networks suffer a loss of local circle structures, whereas NaCl ${\rm H_2O}$-Networks gains more circles. Moreover, in both systems, the size of local circle structures become smaller and smaller with concentration increase. However, for NaCl, these circles have very similar sizes in each individual concentration. On the contrary, in KSCN systems, the difference between these circle structures enlarges with the concentration.

To sum up, as concentration increases, for both O-Networks and ${\rm H_2O}$-Networks, KSCN systems demonstrate much more dramatic variations in their local circle structures. Even though a consistent increase of large-sized local circle structures is observed, sizes of circle structures become more and more diverse. In contrast, NaCl systems show no increase of large-sized circles, instead, a consistent decline of the average size of circle structures is observed. Moreover, sizes of these circles become more and more uniformed.

\section{Conclusion remarks}
For the first time, persistent homology is introduced to quantitatively analyze the intrinsic topological properties of ion aggregation systems and hydrogen-bonding networks. Different from all previous graph and network models, persistent homology analysis does not require a predefined bond length for network construction, and it provides a geometric measurement for inner circle structures.

Two well-studied systems from MD simulations are used to validate our models. It is found that our barcode features capture the essential differences between these two ion aggregation systems very well. The ``ion clusters" and ``extended ion network" models can be quantitatively characterized by our topological measurements. Further, we construct two types of models, i.e., O-Networks and ${\rm H_2O}$-Networks, for analyzing the topological properties of hydrogen-bonding networks. It has been found that, as concentration increases, KSCN systems vary dramatically in their local circle structures. More and more large-sized local circles are generated. In contrast, NaCl systems are much more stable. Their average size of circles declines with concentration. And these circles become more and more uniformed.
Our model provides a whole new perspective in characterizing topological features of structures and can be employed to study complicated network-like systems, including nanomaterials, colloidal systems, biomolecular assemblies, etc.

\section*{Acknowledgments}
The authors would like to thank Dr. Minhaeng Cho and Dr. Jun-Ho Choi for sharing their MD simulation data. This work was supported in part by Nanyang Technological University Startup Grant M4081842.110 and Singapore Ministry of Education Academic Research fund Tier 1 M401110000.

\vspace{0.6cm}


\end{document}